\documentclass[sigconf,screen=true,authordraft=false,review=false,timestamp=false,anonymous=false]{acmart}

\AtBeginDocument{
	\providecommand\BibTeX{{
			\normalfont B\kern-0.5em{\scshape i\kern-0.25em b}\kern-0.8em\TeX}}}

\copyrightyear{2020}
\acmYear{2020}
\setcopyright{acmcopyright}\acmConference[SCF '20]{Symposium on Computational Fabrication}{November 5--6, 2020}{Virtual Event, USA}
\acmBooktitle{Symposium on Computational Fabrication (SCF '20), November 5--6, 2020, Virtual Event, USA}
\acmPrice{15.00}
\acmDOI{10.1145/3424630.3425412}
\acmISBN{978-1-4503-8170-3/20/11}

\usepackage{nicefrac}
\usepackage{booktabs}

\begin{document}
	\newcommand{\norm}[1]{\left\lVert#1\right\rVert}
	\newcommand{\pare}[1]{{\left(#1\right)}}
    \newcommand{\bracket}[1]{{\left[#1\right]}}
    \newcommand{\curly}[1]{{\{#1\}}}
    \newcommand{\rem}[1]{}
    \renewcommand{\quote}[1]{``#1''}
    \newcommand{\cidx}[1]{{}^{\pare{#1}}}
    \newcommand{\connection}{\mathbf{q}}
    \newcommand{\rod}{r}
    \newcommand{\cl}{\mathbf{c}}
    \newcommand{\gc}{{g}}
    \newcommand{\hc}{{h}}
    \newcommand{\var}{\mathbf{v}}
    \newcommand{\realnumbers}{\mathbb{R}}
    \newcommand{\inner}[2]{\langle#1,#2\rangle}
    \newcommand*\diff{\mathop{}\!\mathrm{d}}
    \newcommand{\edge}{\mathbf{e}}
    \newcommand{\vertex}{\mathbf{x}}
    \newcommand{\etal}{{et al.~}}
    \newcommand{\change}[1]{{\color{red}#1}}
    \newcommand{\step}[1]{{\color{teal}#1}}
    \newcommand{\przem}[1]{{\color{blue}#1}}
    \newcommand{\planarangle}{\alpha}
    \newcommand{\w}{\lambda}
    \newcommand{\restl}{l_s}
    \newcommand{\notch}{{N}}
    \newcommand{\notchl}{l}
    \newcommand{\notchul}{\tilde{l}}
    \newcommand{\matlab}{\textsc{Matlab}}
    \newcommand{\tod}[1]{\overline{#1}}
    \newcommand{\cutl}{{\mathcal{L}}}
    \newcommand{\injR}{{ir}}
    \newcommand{\Psur}{\mathcal{P}}
    \newcommand{\Ppla}{\tod{\mathcal{P}}}
    \newcommand{\cladf}{\mathcal{F}} 
    \newcommand{\distf}{\mathcal{C}}

	\title{Design and Fabrication of Elastic Geodesic Grid Structures}

	\author{Stefan Pillwein}
	\affiliation{
		\institution{Technische Universit\"at Wien (TU Wien)}}
	\author{Johanna K\"ubert}
	\affiliation{
		\institution{Technische Universit\"at Wien (TU Wien)}}
    \author{Florian Rist}
	\affiliation{
		\institution{King Abdullah University of Science and Technology (KAUST)}}
	\author{Przemyslaw Musialski}
	\affiliation{
		\institution{New Jersey Institute of Technology (NJIT)}}

	\begin{abstract}
	Elastic geodesic grids (EGG) are lightweight structures that can be easily deployed to approximate designer provided free-form surfaces. 
	In the initial configuration the grids are perfectly flat, during deployment, though, curvature is induced to the structure, as grid elements bend and twist.
	Their layout is found geometrically, it is based on networks of geodesic curves on  free-form design-surfaces.
	Generating a layout with this approach encodes an elasto-kinematic mechanism to the grid that creates the curved shape during deployment.
	In the final state the grid can be fixed to supports and serve for all kinds of purposes like free-form sub-structures, paneling, sun and rain protectors, pavilions, etc.
	
    However, so far these structures have only been investigated using small-scale desktop models. We investigate the scalability of such structures, presenting a medium sized model. It was designed by an architecture student without expert knowledge on elastic structures or differential geometry, just using the elastic geodesic grids design-pipeline.
    We further present a fabrication-process for EGG-models. They can be built quickly and with a small budget. 
	\end{abstract}


	
	\begin{teaserfigure}
		\begin{minipage}[b]{0.47\textwidth}
			\centering
			\includegraphics[width=0.84\textwidth]{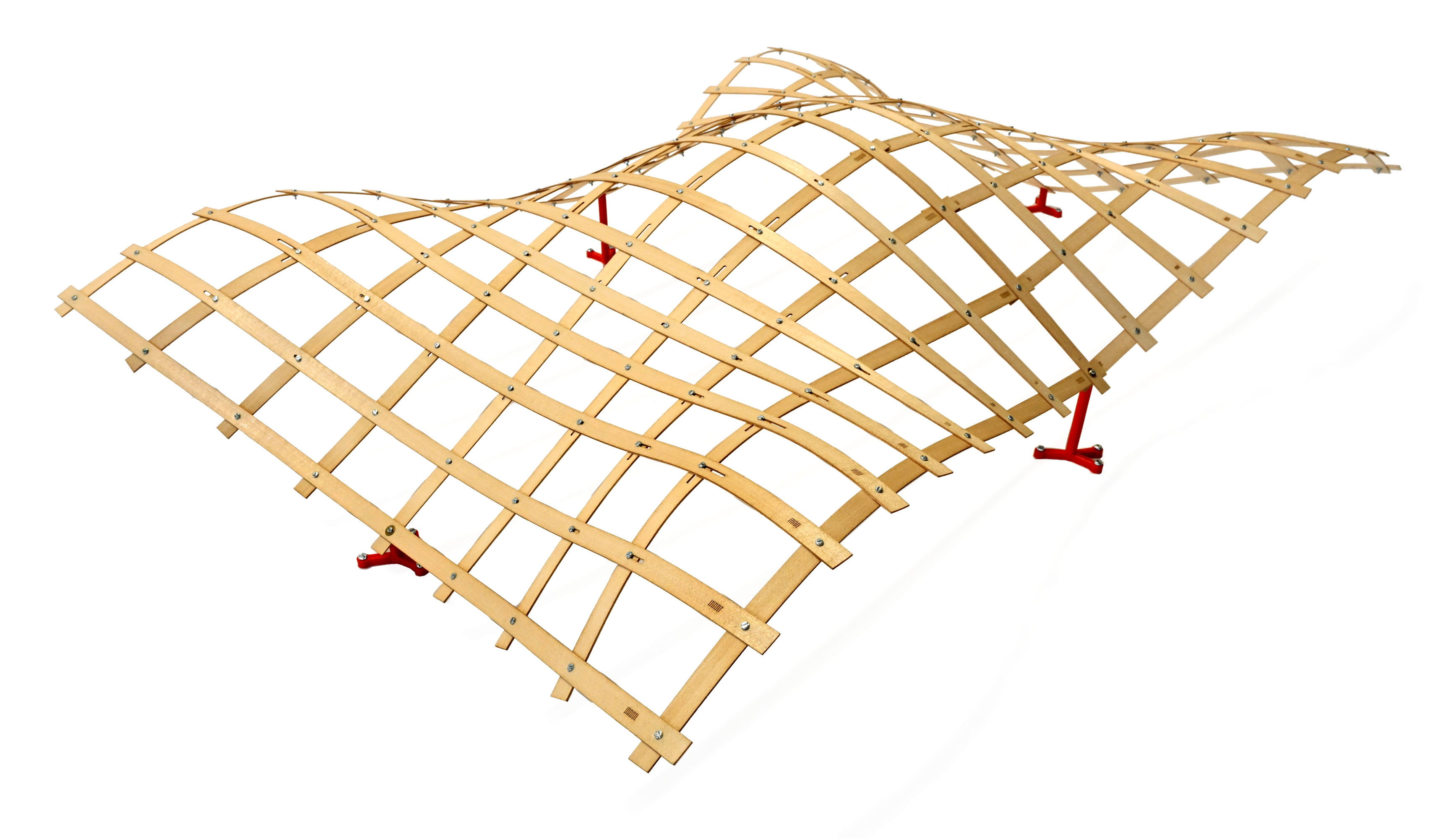}
		\end{minipage}
		\hfill
		\begin{minipage}[b]{0.47\textwidth}
			\centering
			\includegraphics[width=0.84\textwidth]{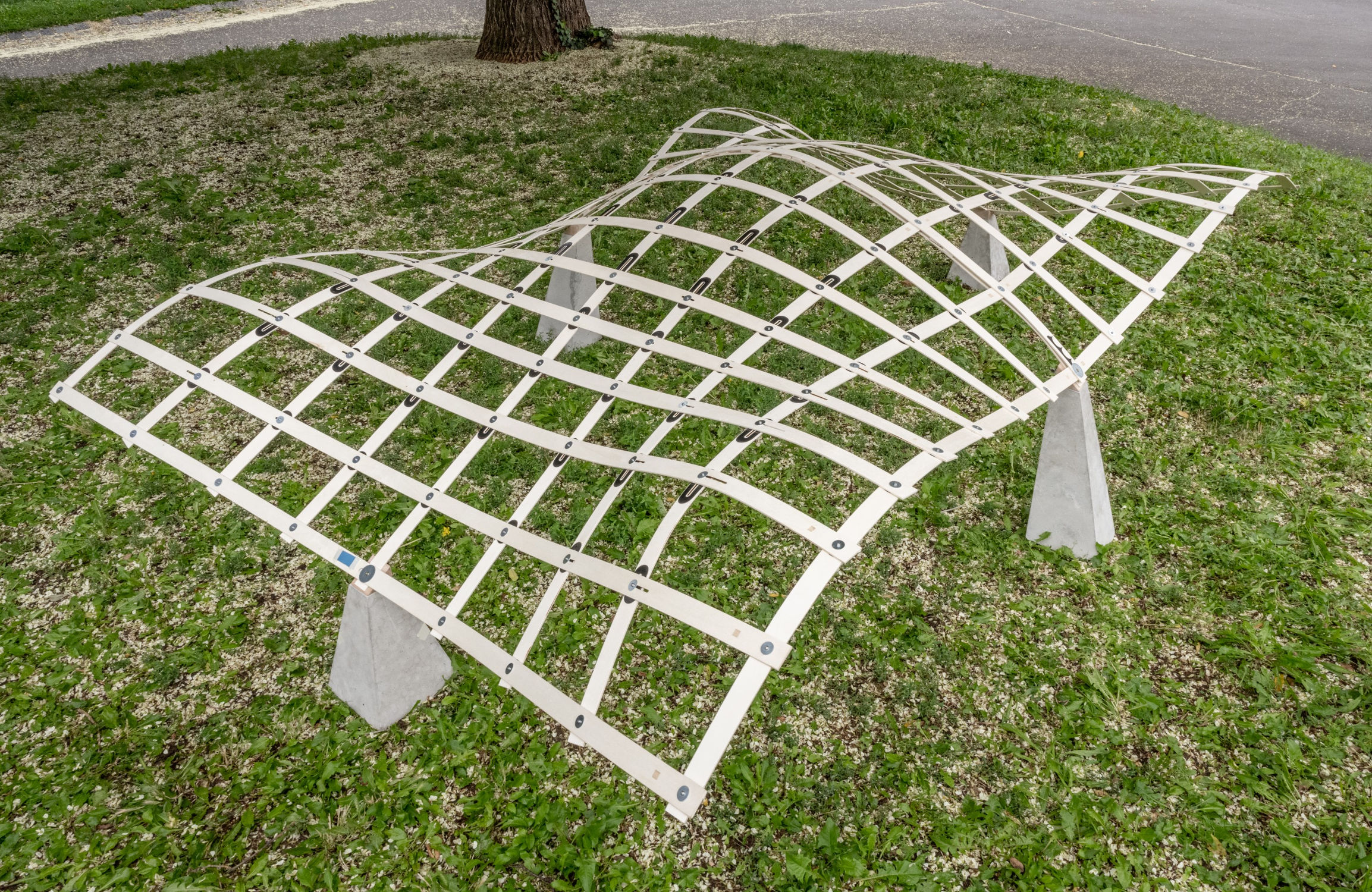}
		\end{minipage}
		\caption{We present a medium sized structure that was built to investigate the scalability of the elastic geodesic grids approach. The desktop model to the left was scaled by a factor of four to receive a structure of $3.1 \times 2.1 \times 0.9$ meters in size.}
		\label{fig:grids}
		\vspace{0.2 cm}
	\end{teaserfigure}

	\maketitle

	\section{Introduction}\label{sec:concept}
    We are surrounded by curved elements and structures, they are aesthetically pleasing and functional, however, they are not always easy to produce. 
    The curved shape of the structure in Figure \ref{fig:grids} was generated by deploying a planar grid of initially straight, elastic lamellas.
    Elastic geodesic grids (EGG), proposed by \cite{Pillwein2020}, can be deployed from a planar state to a spatial state, where the spatial state approximates a design surface. 
    The transformation between these states can be performed by simply pulling two corners of the planar grid apart and applying some additional bending. 
    The mechanism that causes the grid to take a spatial shape is simple: 
    The lamellas in the planar grid are not parallel, which makes it rigid in the plane. However, it is not rigid in space and, when actuated, the grid buckles to a spatial shape.
    
    The deployment approach makes the creation of doubly-curved shapes quick and material efficient.
    Lightness and easy assembly make the grids applicable for mobile purposes. Also, their fabrication does not involve complicated techniques, the presented structure was built using simple resources like plywood, screws, washers and nuts. If advanced tools are not available, it could even be built using a tape measure, a drill, and a saw.
    Light grid structures are particularly suitable for applications where large spans and low weight are required. This makes them potential candidates for architectural purposes.
    
    The scope of this paper is to analyze how elastic geodesic grid models can be implemented on a larger scale, and how the original approach to the design and fabrication of these structures can be improved. 
    \begin{figure}
	\centering
	\includegraphics[width=\columnwidth]{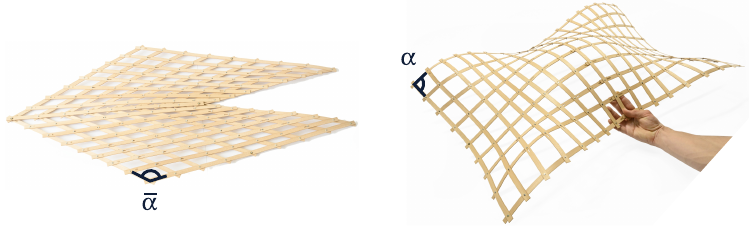}
	\caption{An elastic geodesic grid in the planar and the deployed configuration, the deployment is based on $ \tod{\alpha} \rightarrow \alpha$.}
	\label{fig:pla_dep}
    \end{figure}
    It features the following contributions:
    \begin{itemize}
        \item The scalability and fabrication of the EGG-approach is investigated, and a medium sized model is built.
        \item A geometric solution to the smoothing-problem in the approach of \cite{Pillwein2020} is proposed by introducing multiple modules, i.e. by decomposing the input surface.
        \item The negative effects of friction in the physical grid during deployment are analyzed and systematically minimized.
    \end{itemize}
    In Section \ref{sec:related_work}, the related work is reviewed, Section \ref{sec:egg}  reviews the EGG-approach, Section \ref{sec:challenges} examines fabrication and design challenges, and presents solutions. A medium sized model and the fabrication process are presented in Section \ref{sec:pavilion}, and finally, Section \ref{sec:discussion} features discussion and conclusions.

    \section{Related Work}\label{sec:related_work}
    \paragraph{Active-Bending Paradigm}
    In the computer science, architecture and engineering   communities, the \textit{active bending} paradigm \cite{Lienhard2013} and easy-to-use computational methods  have sparked a new wave of interest in elastic structures that are deformed to create curved shapes.
    Until recent advances in computing elastic structures fast and physically accurate, they could only be form-found empirically \cite{gengnagel2013active}. 
  
    \paragraph{Gridshells and Deployment-Approaches}
    A lot of research is currently being carried out on gridshell-structures that can be deployed. They can be classified based on their deployment mechanism: inscribed in a grid that is deployed \cite{Panetta2019,X_shells_pavilion,Soriano2019,Pillwein2020} or by other external mechanisms like inflatable air cushions \cite{pneumatic}.
    We are interested in the first case. The design approaches of inscribing the deployment mechanism into the grid, however, differ a lot: 
    
    To create an X-Shell \cite{Panetta2019}, a planar grid layout is designed and actuated with a physical simulator. The grid curves of the planar layout do not have to be straight. In multiple layout iterations the designer finds a satisfactory shape by changing the planar design. This approach does not depend on a target surface. Target surfaces, however, can be approximated if a good planar initialization is provided by the designer. In this case, a grid that approximates the target surface closely can be found using shape optimization.
    The bending and twisting behaviour of the rods can be controlled by using different shapes of cross sections. This has a direct effect on the shape of the deployed X-Shell, as the authors show.
    \cite{X_shells_pavilion} investigate X-Shells  hands-on with the construction of a pavilion of GFRP-rods, measuring 3.2 × 3.2 × 3.6 meters.
    
    In the G-Shells-approach \cite{Soriano2019} propose to planarize a specific geodesic grid using physical simulation inside an evolutionary multi-objective solver. This way, a geometric error is introduced, so the flat grid cannot be deployed to match the geodesic grid perfectly, but it can come close, creating beautiful shapes. G-Shells look similar to elastic geodesic grids, as they are also built from thin lamellas.
    
    The EGG-approach of \cite{Pillwein2020} takes a design surface as  input and results in a grid layout that can be deployed to approximate this surface. The layout is found geometrically and simultaneously on the design surface and in the plane. This makes the computation of grid layouts efficient, as neither design iterations nor physical simulation are needed for finding layouts.
    Despite the similarity of using geodesics as grid curves, in contrast to \cite{Soriano2019}, elastic geodesic grids use the concept of notches. This prevents introducing a geometric error in the grid layout, but also makes the deployment process more complicated as sliding of members is necessary.
    However, this method does not feature arbitrary curve networks as it is limited to geodesics as grid curves. This poses a restriction on the space of solutions of deployable grids using an inscribed deployment mechanism. Besides this limitation, design surfaces with high Gaussian curvature might need to be smoothed before a suitable grid layout is feasible.
    
    \cite{Schling2018} propose structures based on networks of asymptotic curves on a surface, which are found via optimization using the guided projection scheme. These structures can be assembled in segments that are initially flat, but transform to a curved state by their internal forces.
    
    \paragraph{Other Deployable Surfaces}
    Besides the architecture domain, there has been also extensive research in the computer graphics literature on various methods for deployable structures. 
    One more way to easily construct spatial shapes from flat sheets is by appropriately folding paper \cite{Mitani2004b,Massarwi2007a}, which is inherently related to the Japanese art of Origami \cite{Dudte2016}. 
    Other works deal with curved folds and efficient actuation of spatial objects from flat sheets \cite{Kilian2008,Kilian2017}. 
    Elastic geodesic grids are related to these approaches in terms of being deployable from a planar initial state, however, the main difference is that they are elastic and approximate doubly-curved surfaces.  

    In fact, a lot of attention has been paid to the design of doubly-curved surfaces which can be deployed from planar configurations due to the ease of fabrication. One way of achieving this goal is by using auxetic materials \cite{Konakovic2016} which can nestle to doubly-curved spatial objects, or in combination with appropriate actuation techniques, can be used to construct complex spatial objects \cite{Konakovic-Lukovic2018}. The main difference to the elastic geodesic grids method of Pillwein et al.~\shortcite{Pillwein2020} is that these structures do not use elastic bending to reach the spatial shape. 

    Also the idea of storing energy in a planar configuration in order to approximate spatial shapes has been explored. This can be done, for instance, by using prestressed latex membranes in order to actuate planar structures into free-form shapes \cite{Guseinov2017}, or to predefine flexible micro-structures which deform to desired shapes if they are combined and interact \cite{Malomo2018a}. A combination of flexible rods and prestressed membranes lead to Kirchhoff-Plateau surfaces that allow easy planar fabrication and deployment \cite{Perez2017a}. 
    Thus, the elastic geodesic grids approach is based on the assumption that the elastic elements can bend and twist, but not stretch, and must therefore maintain the same length in the planar as well as in the spatial configuration.

    \section{Elastic Geodesic Grids (EGG)}\label{sec:egg}
    The choice of using geodesics as grid curves is mainly motivated by practical reasons: A thin strip of material can be wrapped on a surface and interpreted as a tangential strip with the centerline of the strip being a geodesic on the surface.
    This means that the planar grid can be fabricated from straight lamellas, where the deployment-mechanism encoded in the layout does the wrapping.
    Fabricating an elastic grid with straight lamellas is material and cost efficient, allows easy assembly, and facilitates transport, as the lamellas can be stacked in a small volume.
    
    Besides practicability, using geodesics as grid curves is also motivated by geometry and physics.
    Pillwein etal. \shortcite{Pillwein2020} show, if the mechanics of the grid strongly correlate with the geometric properties of geodesics, the shape of the deployed grid will match the shape of the initial geodesic grid closely. 
    In the following we briefly recall the approach of EGG. 
    
    \subsection{Geometric Background}   \label{sec:background}
    Geodesics have no geodesic curvature, their curvature is the normal curvature of the surface. Their torsion is the geodesic torsion of the surface.
    For the elements of the physical grid this implies: 
    Bending should be easy around one axis, impossible around the second axis, twisting should be easy as well, and stretching should be impossible. This behaviour is inherent to thin lamellas made from non-stretching material.
    At points where two geodesics meet, their principal normals coincide with the surface normal. 
    To comply with this characteristic, lamellas are connected by screws. 
    During deployment, the screws act as principal normals of the centerlines at the connections.

    The purpose of an elastic geodesic grid is to approximate a design surface in the deployed state, hence, the shape of this surface needs to be encoded in the grid layout.
    Therefore, a layout is computed by finding a planar grid and a geodesic grid simultaneously, which must meet the following main conditions:
    \renewcommand{\theenumi}{\roman{enumi}}
    \begin{enumerate}
        \item Total lengths of grid curves on the surface and in the plane have to be  equal. \label{1}
        \item Partial lengths between connections on the four boundary members have to be equal. \label{2}
    \end{enumerate}
    These conditions are based on the following idea: 
    If all lengths between connections were equal in both states and the planar physical grid is deployed, it will lie on an isometry of the design surface. To make it lie on the design surface, it just needs to be bent.
    
    However, by complying with (\ref{1}) and (\ref{2}), lengths between inner connections of the grid will not match for arbitrary surfaces. This is known in differential geometry \cite{Lagally}: In general, a geodesic grid cannot be planarized by only changing the angles between grid curves.
    The geometric error, caused by having non matching lengths between inner connections, would impair the quality of the approximation.
    To avoid this problem, \cite{Pillwein2020} introduced sliding notches at inner connections of the grid.
    They allow for a certain amount of sliding and can be implemented physically by elongated holes.
    
    The EGG-approach intertwines geometry and physics to receive curved grids  that approximate design surfaces. The shape of these grids is partly driven by geometric constraints and also by the stiffness of the lamellas with respect to bending, twisting and stretching.

    \subsection{Computation of  EGG-Layouts} \label{sec:approach}
    The computational aspects of the design process proposed by \cite{Pillwein2020} will be briefly summarized below.
    
    \begin{figure}[t]
	\centering
	\includegraphics[width=\columnwidth]{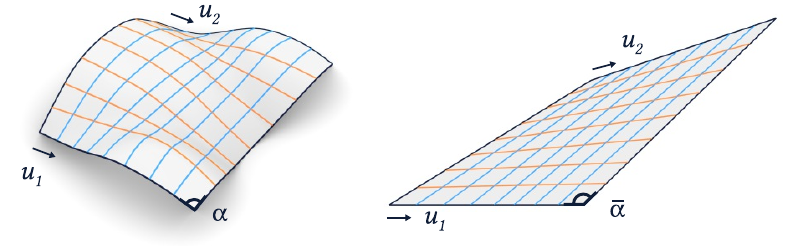}
	\vspace{-20pt}
	\caption{A surface patch and a planar patch carrying two families of grid curves. }
	\label{fig:layout}
    \end{figure}
    
    The design of an elastic grid is initialized by defining a surface patch. It is the part of the design-surface that will actually be approximated by the grid. To frame the patch, four boundary geodesics have to be chosen that form a convex geodesic quad.
    
    As mentioned earlier, the planar grid and the grid on the surface are computed simultaneously. Hence, the next step is to initialize the planar configuration. The planar patch is defined by a planar quad with straight edges and the same side lengths as the geodesic quad. This planar quad has one degree of freedom, as one angle can be chosen freely to fix its shape uniquely. 
    
    As already mentioned in Section \ref{sec:background}, a grid layout needs to comply with the conditions (\ref{1}) and (\ref{2}). The coordinates on boundary members for the planar and the spatial configuration are coupled by condition (\ref{2}). Figure \ref{fig:layout} illustrates the coupled coordinates $u_1$ and $u_2$ for the blue family of members. Please note that the boundary members on the surface and in the plane are parametrized with the same constant speed.
    By choosing start and end coordinates $(u_1,u_2)$, a blue geodesic curve on the surface and in the plane can be defined. However, to obey condition (\ref{1}), the lengths of these two curves need to be equal. This corresponds to the inability of lamellas to stretch or shrink.
    
    Finding a single grid curve maintaining (\ref{1}) and (\ref{2}) is quite simple: If we look at the setup in Figure \ref{fig:layout}, starting with some $\overline{\alpha}$ and a combination of $(u_1,u_2)$ coordinates, $\overline{\alpha}$ and $(u_1,u_2)$ can be varied until the lengths of the geodesic on the surface and in the plane match.
    
    However, to find a global solution and cover the whole surface patch with valid grid curves \cite{Pillwein2020} follow a different, more systematic approach.
    For both families of grid curves this is done in the following way:
    
    First, distance maps are computed for the surface patch and the planar patch. 
    A distance map is a function that assigns the geodesic distance to every combination of coordinates on opposite boundaries. In practice, they are computed by just evaluating a finite number of coordinate-samples, values between the samples are interpolated linearly.
    For the blue family in Figure \ref{fig:layout}, two distance maps, one for the planar and one for the  spatial configuration, need to be computed. The distance map for the surface patch $\mathcal{D}(u_1,u_2)$ is just a function of the boundary coordinates $u_1$ and $u_2$. The planar distance map $\overline{\mathcal{D}}(u_1,u_2,\overline{\alpha})$ is also a function of the angle $\overline{\alpha}$ that fixes the shape in the planar quad. This is due to the fact that the shape of the planar quad influences inner distances.
    If two distance maps $\mathcal{D}, \overline{\mathcal{D}}$ are intersected, the common points describe geodesics on the surface and in the plane that obey (\ref{1}) and (\ref{2}).
    
    There is one degree of freedom to find good-quality grids: the angle $\overline{\alpha}$.
    Good choices of this angle deliver a compact planar layout and allow smooth deployment of the grid. However, poor choices result in extra crossings of grid curves or introduce triangles to the grid. This would destroy the kinematic mechanism, making deployment impossible.
    
    In an optimization procedure a feasible domain for the angle $\overline{\alpha}$ is computed, giving the designer a certain amount of freedom to define the shape of the planar quad. The procedure results in cladding functions that describe the cladding of the surface patch with two families of geodesics.
     
    The layout of a grid can be defined by just picking points on the cladding functions. To capture surface features like bumps well, \cite{Pillwein2020} propose an automatic layout strategy. Please refer to the paper for details.
    
    After having defined the layout, the grid and supports for mounting can be fabricated, the grid can be deployed and fixed to the supports to produce the desired shape. 
    
    \paragraph{Feasibility of Surface Patches}
    After having defined a surface patch, its feasibility for the EGG-process can be checked by a simple geometric criterion \cite{Pillwein2020}: 
    \begin{equation*}
        (e-\overline{e}) (f-\overline{f}) < 0~,
    \end{equation*}
    where $e,f$ are the lengths of the geodesic diagonals on the surface patch, and $\overline{e}, \overline{f}$ are the lengths of the diagonals in the planar patch.
    
    Besides this restriction on the shape of the patch, high Gaussian curvature $K$ can cause problems for the quality of the approximation. Smoothing might be needed to reduce $K$ before the patch can be used. This will be discussed more closely in Section \ref{sec:modularity}.
    
   \begin{figure}[t]
    	\centering
    	\includegraphics[width=\columnwidth]{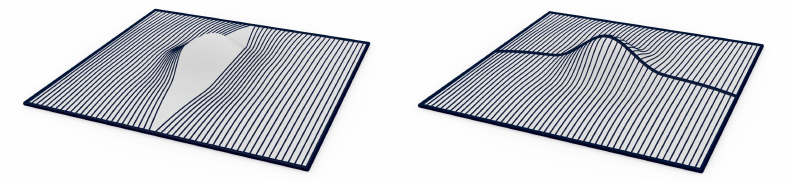}
    
    	\caption{Splitting surface patches allows an even coverage with geodesics and makes smoothing redundant. Left: Shortest geodesics between boundaries show problems caused by non uniqueness. Right: Splitting the surface allows an even coverage of both patches.}
    	\label{fig:unique}
    \end{figure}

    \section{Fabrication Challenges}\label{sec:challenges}

    There are some challenges associated with the design and the implementation of elastic geodesic grids on different scales. This section presents some of the challenges and possible solutions. The medium-scale model shown in Figures \ref{fig:grids}, \ref{fig:side} and \ref{fig:pavilion} was used to evaluate these approaches. The measures presented are intended to help improve the general feasibility of the approach for larger models.
    
    \subsection{Geometric Limitations and Modularity} \label{sec:modularity}
    Choosing geodesic curves as grid curves poses some restrictions on the class of design surfaces that can be approximated.
    As already mentioned earlier, high Gaussian curvature can cause problems as a surface patch may have to be smoothed before it can be used. From a designer-perspective this is undesirable, so the problem will be analyzed and a solution will be proposed.
    
    The problem is linked to the use of shortest geodesics as grid curves. This class of geodesics can be defined as shortest connections of two points on a surface. Here, the points are situated on opposite boundaries of the surface patch (cf. Figure \ref{fig:layout}). It may happen that these points can be connected by more than one shortest geodesic. In other words, shortest geodesics may not be unique.
    This property is problematic, because it implicates that some areas of the surface cannot be covered with shortest geodesics, and therefore cannot be approximated properly.
    However, the uniqueness can be checked, because a link with the Gaussian curvature $K$ is known \cite{Carmo1992}: 
    \begin{equation} 
    \label{eq:injectivity} 
    \injR(p) \ge \frac{\pi}{\sqrt{K_\text{max}}} \,.
    \end{equation}
    The left side of the inequality $\injR(p)$ is the injectivity radius for geodesic curves for each surface point $p$ and $K_\text{max}$ is maximum of the Gaussian curvature of the design surface.
    In essence, if all geodesics on the surface patch are shorter than the value on the right hand side of Expression (\ref{eq:injectivity}), problems with non uniqueness of shortest geodesics do not exist.
    The approach of \cite{Pillwein2020} is to smooth the surface, i.e. to reduce K to eliminate non unique shortest geodesics.
    Although smoothing enables the computation of a functional grid, it will most likely interfere with the design intent.
    
    However, smaller surface patches tend not to need smoothing, because they are less prone to feature non unique shortest geodesics. Geodesics will become shorter without reducing K, and inequality (\ref{eq:injectivity}) is more likely to be fulfilled.
    From this insight it becomes clear that splitting the surface patch is a reasonable strategy to omit smoothing.

    Clever splitting of the surface patch can reduce the problem further: Expression (\ref{eq:injectivity}) does not capture the location of the peak on the surface patch, but only delivers a quick check. 
    The most effective way to get rid of problems with non unique shortest geodesics is to place the boundaries of the modules directly over the peaks on the surface patch.
    This can be illustrated with a simple example, shown in Figure \ref{fig:unique}. The central peak causes shortest geodesics between the boundaries which are not unique. 
    There is no shortest geodesic which goes over the peak.
    The part of the surface that is not covered can not be encoded in the grid properly. If the surface is split into two modules, where one  geodesic boundary leads directly over the peak, shortest geodesics become unique, although Expression (\ref{eq:injectivity}) might still indicate otherwise, as it provides a sufficient but not necessary condition for uniqueness.
    Please note that splitting must finally result in modules with shortest geodesics as boundaries.

    From a practical point of view, the deployment of a single large-scale elastic grid is certainly not a trivial task because of handling issues.
    Splitting up a single surface patch into smaller patches, which allows the fabrication of several separate grids that can be deployed and combined sequentially, simplifies this task.
    In the physical implementation grids can be simply connected along their common lamella, as can be seen in Figure \ref{fig:small_scale}. 
        Please note that there are no notches on boundary members, because of (\ref{2}).
    
    As proposed by \cite{Pillwein2020}, supports are placed at connections on the boundary curves close to their extrema of curvature. When splitting up the surface patch into modules, a support can thus be shared by two or more grids.
    Eliminating inner supports and only keeping supports on the external boundaries is desirable, however, it would require an analysis of the forces arriving at supports. If they cancel out or have a small residual, the individual support can be omitted.
    
    \begin{figure}
	\centering
	\includegraphics[width=\columnwidth]{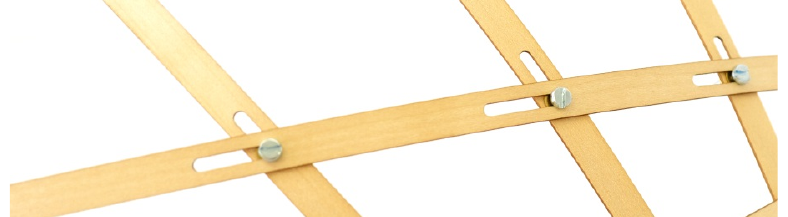}
	\caption{Notches at the connection of grid members. They enable partial sliding, which is necessary for the close approximation of the design surface. They can be implemented by elongating holes.}
	\label{fig:notch}
    \end{figure}
    
    \subsection{Notches and Friction} \label{sec:friction}
    The quality of the approximation strongly depends on the concept of notches, as outlined in Section \ref{sec:background}. Notches enable grid members to slide for a predefined length and direction at their connections. Figure \ref{fig:notch} shows notches in a physical grid.

    While sliding of members in the notches delivers perfect results in simulations, friction poses a problem for the sliding process of physical grids. Lamellas often get stuck in the notches, because forces that are supposed to push the lamellas in the right direction inside the notches are counteracted by friction.
    By interfering with sliding, the quality of the approximation suffers, and local stresses arise as some lamellas buckle. Also, the dissipation of energy by friction increases the force necessary to deploy the grid.
    
    The regions that contribute to the overall friction during deployment are the contact areas of lamellas, the screw threads and the notches, and the lamellas and the washers. Table \ref{tab:mu} summarizes empirically determined friction coefficients. The goals of easy deployment on one hand and secure final fixing on the other hand can be fulfilled by reducing the friction between the lamellas, but maintaining friction between lamellas and washers.
    
    \begin{figure}[t]
    		\begin{minipage}[b]{0.22\textwidth}
    			\centering
    			\includegraphics[width=\textwidth]{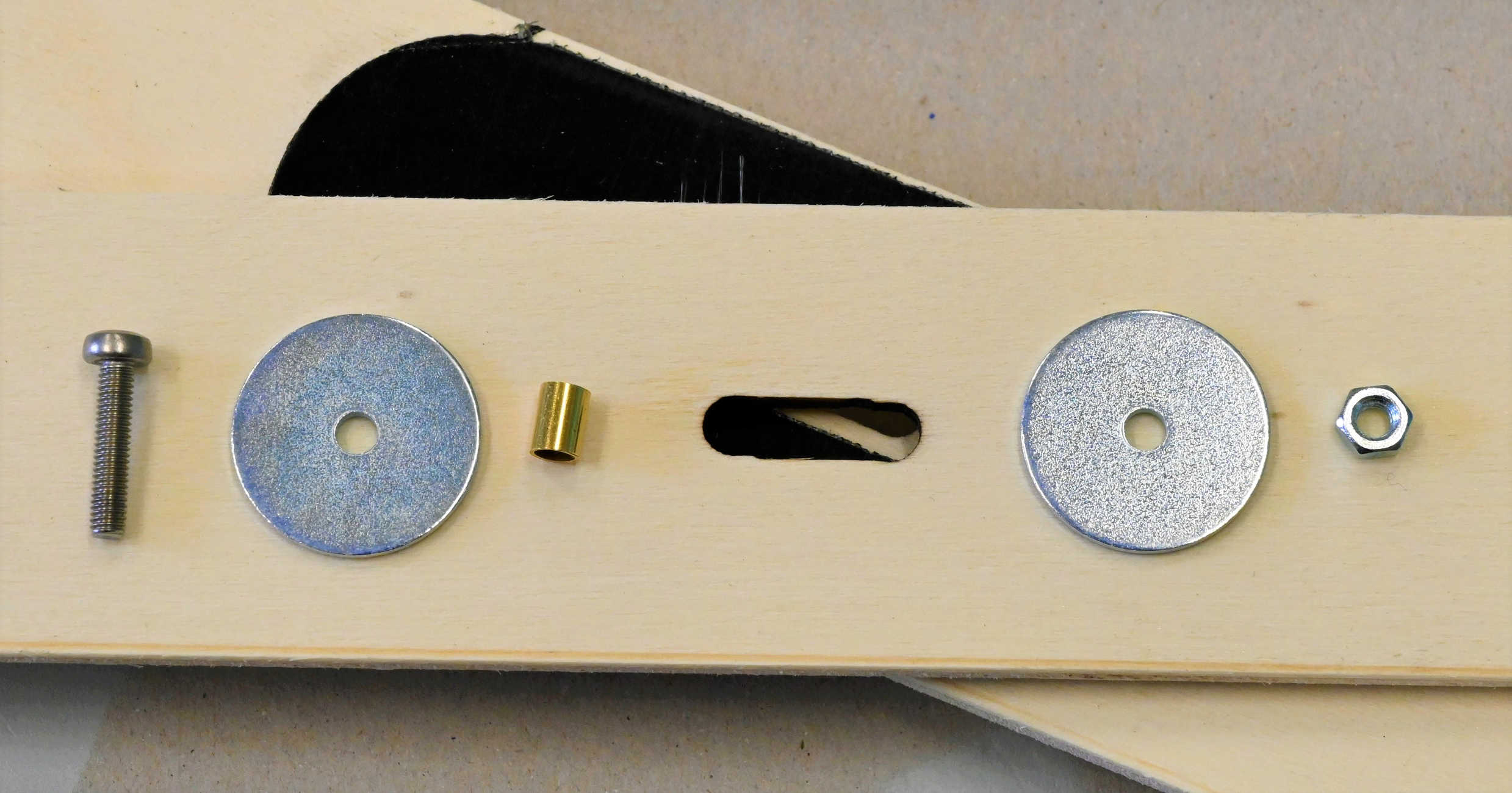}
    		\end{minipage}
    		\hfill
    		\begin{minipage}[b]{0.22\textwidth}
    			\centering
    			\includegraphics[width=\textwidth]{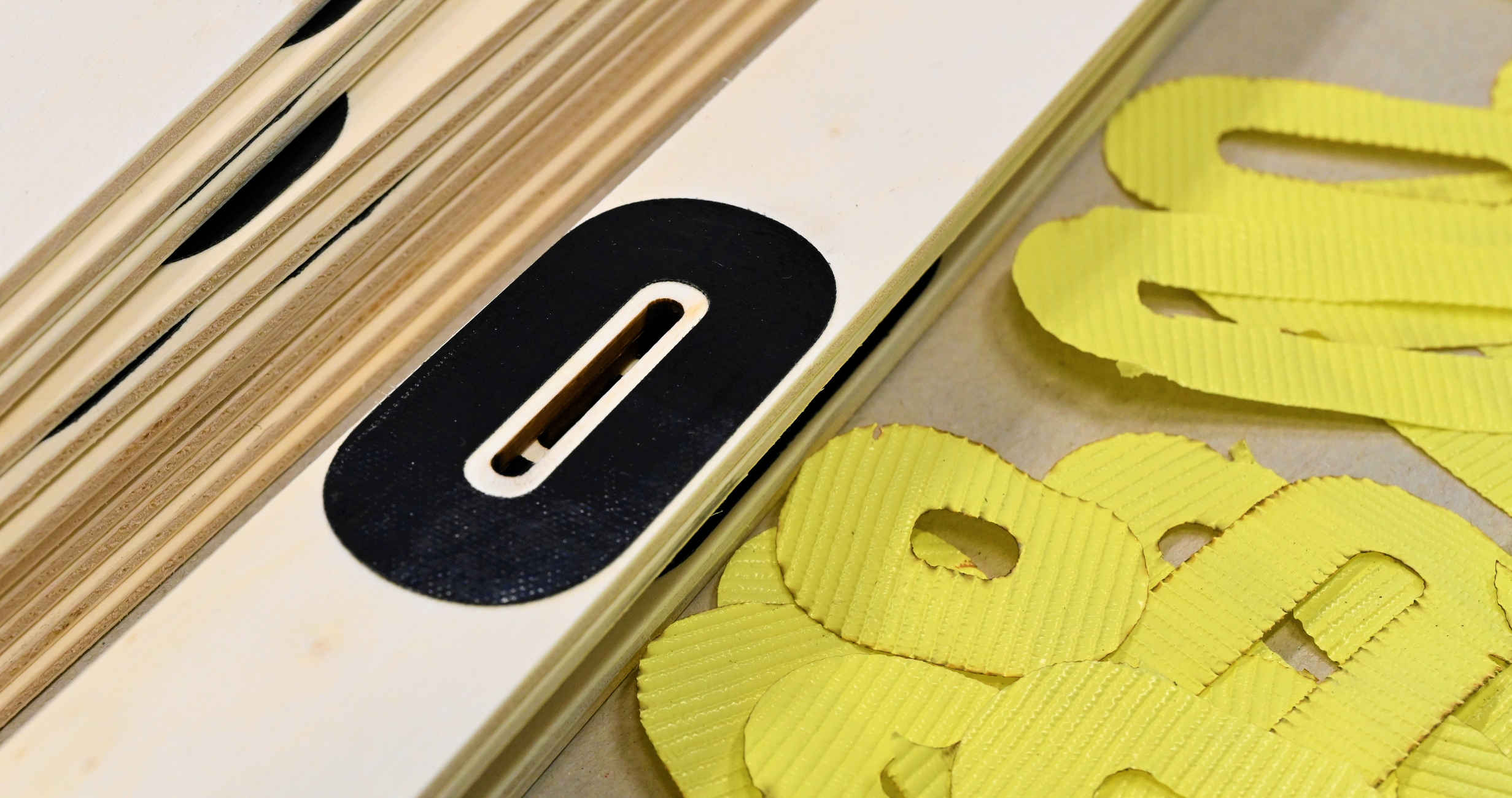}
    		\end{minipage}
    		\caption{Design features to minimize friction. Left: Contact of the screw thread and the notch wall is prohibited by a brass shell. Right: PTFE-foil stickers reduce friction at lamella-to-lamella contact areas.}
    		\label{fig:friction}
    		\vspace{0.2 cm}
    \end{figure}
    
    An effective measure to reduce friction is to equip the lamella to lamella contact areas with a PTFE-foil layer. The friction coefficient of PTFE is quite low, and it is robust. For every notch an individual PTFE-sticker was laser-cut from an adhesive foil. The stickers cover the immediate contact area for every notch, so there is no contact of lamellas. 
    Brass shells furthermore reduce the friction of the screw in the notch. Figure \ref{fig:friction} shows the measures taken to reduce friction.

     \begin{table}[b]
     \caption{Empirically determined friction coefficients $\mu$ for static and sliding friction; the evaluated plywood is poplar.}
        \begin{tabular}{cccc}
           Material &    ~~~~~~~~~~~~$\mu_{static}$ ~~~~~~~~~~~~        &  ~~~~~~~~~~~~ $\mu_{sliding}$~~~~~~~~~~~~\\[0.02cm]
        \midrule
       Plywood-Plywood   & 0.436                         & 0.273                  \\
        PTFE-PTFE       & 0.163                         & 0.091                   \\
        Plywood-Steel    & 0.252                         & 0.203                   \\[0.02cm]
        
        \end{tabular}
        
        \label{tab:mu}
    \end{table} 
    
    \subsection{Scaling of the Structure}\label{sec:scaling}
    
    Fabricating small prototypical models to investigate the qualities of the design is common practice in architecture. Once a suitable material is found, designing EGG models on the same scale is easy. 
    Analyzing the models of \cite{Pillwein2020}, we provide the rough empiric rule that a lamella-width of around two percent of the desired output size works well. The lamella-height follows from the considerations in Section \ref{sec:background} as smaller than $\nicefrac{1}{5}$ of the width.
    However, building EGG models on a larger scale is not straightforward, as we will investigate in this section. 
    
    The shape of a deployed elastic grid is driven by the stiffness-parameters of the lamellas. The related approach of \cite{Panetta2019} even uses different shapes of cross sections, which induce different stiffness-parameters to influence the shape of a deployed X-Shell. 
    The stiffness of an element like a lamella can be expressed with regard to some kind of elastic deformation. It determines the amount of stress which is induced by this deformation. The magnitude of a stiffness-parameter is set by the geometry of the cross section and by the material parameters. 
    In our case, when wrapping a lamella on the surface patch, curvature is prescribed. Knowing the bending stiffness, the internal stresses due to bending can be computed.
    
    To get an insight into the matter, we will consider linear scaling of the geometry, which also means linear scaling of the cross sections of the lamellas.
    
    Let us assume that the lamellas of the deployed grid agree well with the scaled surface patch. This enables us to estimate the curvature of the lamellas and to compute the stresses due to bending w.r.t. the scaling factor $f$:
    \begin{align*} 
    \sigma_{B,max}\scriptstyle(f) = &  \pm \frac{M\scriptstyle(f)}{I\scriptstyle(f)}  \frac{hf}{2}= \pm \frac{E~ I\scriptstyle(f) \frac{\kappa}{f} }{I\scriptstyle(f)}  \frac{hf}{2} =   \pm \frac{E \kappa  h}{2} \,,
    \end{align*}
    where $\sigma_{B,max}$ is the maximum normal-stress induced by bending, $M$ is the bending-moment, $I$ is the moment of inertia, and $\kappa$ is the curvature.
    This means that the bending stresses are constant under linear scaling, as the curvature decreases at the same rate as the height of the cross section increases.
    
\begin{figure}[t]
    	\centering
    	\includegraphics[width=\columnwidth]{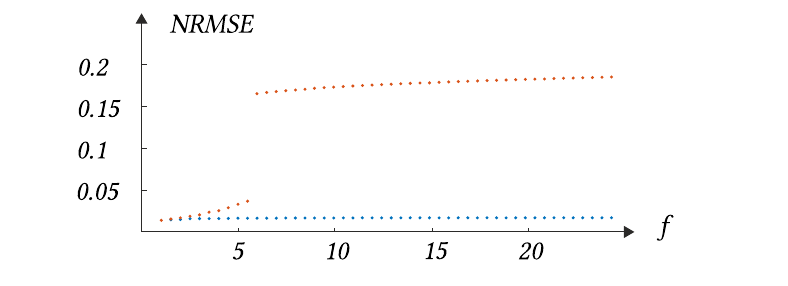}
    	\vspace{-22pt}
    	\caption{The effects of scaling an EGG. The NRMS-Error is used to measure the deviation of scaled simulated grids to the design surface.
    	Blue dots represent simulation results, if gravity is neglected, orange dots represent results including gravity. The jump represents the collapse of the structure under its own weight.}
    	\label{fig:scaling_f}
    \end{figure}

    The shape of the scaled model is determined by internal stresses, mainly caused by bending and self-weight. As the scaling factor $f$ increases, stresses caused by self-weight grow significantly, whereas stresses caused by bending remain constant.
    The strong increase of stresses caused by self-weight is due to the fact that it grows cubically w.r.t. the scaling factor $f$: 
    \begin{align*} 
    F_G\scriptstyle(f) =  &  m\scriptstyle(f) \displaystyle g = V\scriptstyle(f) \displaystyle  \rho  g =  h  b  l  f^3  \rho  g \,.
    \end{align*}
    Figure \ref{fig:scaling_f} shows numerical results of the impact of scaling on the shape, generated with a physical simulation using the Discrete Elastic Rods model \cite{Bergou2008, Bergou2010}. The implementation we used is based on the implementation by \cite{Vekhter2019}, which features the simulation of self-weight.
    To compare the shapes of linearly scaled versions of the grid in Figure \ref{fig:pla_dep}, the Normalized Root Mean Square Error ($NRMSE$) was used. The difference between the predicted and the observed values in the $NRMSE$ are the distances $d_i$ between points on the centerlines of a simulated, deployed grid and their nearest neighbor in the surface patch. The inverse scale factor acts as the normalization factor: 
    \begin{align*} 
    NRMSE = \frac{1}{f}\sqrt{ \frac{1}{n} \displaystyle \sum_{i=1}^{n} d_i^2  } ~.
    \end{align*}
    Figure \ref{fig:scaling_f} shows that as long as gravity is neglected, linear scaling does not influence the shape substantially. However, taking gravity into account, the cubical increase of gravitational loads makes linear scaling feasible only for small scale factors. 
    Figure \ref{fig:deviations} shows the simulated shape of the desktop model, it corresponds to the leftmost data point in Figure \ref{fig:scaling_f} ($f=1$).
    
    \paragraph{Scaling Strategies}
    To enable the construction of larger structures, the gravitational forces somehow need to be kept small, while we expect a positive effect of increasing the relevance of bending stresses.
    There are two options to tackle the problem: either scale non uniformly or change the  material.
    
    Non uniform scaling can be applied, but only inside certain width-to-height bounds of the cross section, as mentioned in Section \ref{sec:background}. Increasing $h$ without keeping a high width-to-height ratio will corrupt the EGG design concept, and thus worsen the quality of the approximation.

\begin{figure}[t]
    	\centering
    	\includegraphics[width=\columnwidth]{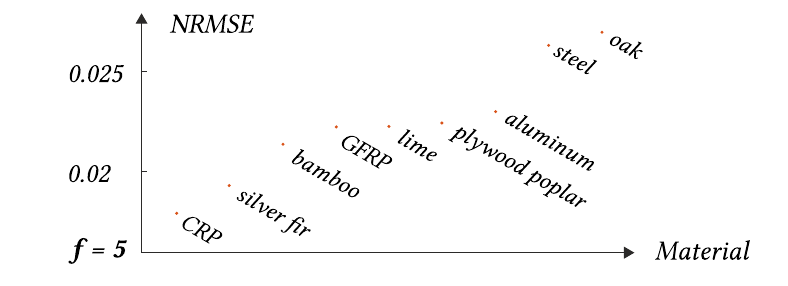}
    	\vspace{-22pt}
    	\caption{The performance of different materials at $f=5$. }
    	\label{fig:materials}
    \end{figure}

\begin{figure}[b]
    	\centering
    	\includegraphics[width=0.95\columnwidth]{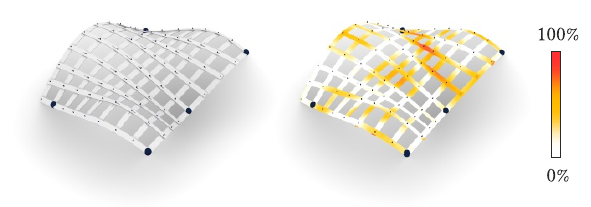}
    	\vspace{-10pt}
    	\caption{Simulated elastic grid and deviations of its shape to the shape of the surface patch. The deviations are colored and relate to the Euclidean distance w.r.t. the width of a lamella.     	
    	The mean deviation is 2 mm for absolute dimensions of 0.43 x 0.57 x 0.1 m}
    	\label{fig:deviations}
    \end{figure}

    Changing the material, however, allows improvement in two ways, namely by changing the elastic modulus $E$ and the density $\rho$. The elastic modulus only influences bending stresses, and the density only influences the gravitational forces. The most practical approach to scaling an EGG is looking for a material with the right ratio of the two.
    In fact, this ratio is called specific modulus $\lambda = \nicefrac{E}{\rho}$ and is well known in light weight engineering like aerospace design. There it is used for parts whose shape is driven by stiffness, like the wings. Figure \ref{fig:materials} shows the impact of using different materials on the shape of an EGG for a scaling-factor of $f=5$.

    However, there is a further restriction that needs to be considered. When increasing the elastic modulus, the stresses due to bending  grow proportionally, as the curvature is prescribed.
    So it needs to be checked that the stresses due to bending do not exceed the strength of the material, as the structure would break otherwise.
    
    To summarize, properly scaling the grid by a factor $f$ can be achieved by tuning the specific modulus $\lambda$ to receive the best possible shape. In practice, this suggests that the bigger an EGG gets the more high-quality materials need to be used. Additionally, $\sigma_{B,max}$ and the strength of the material need to be checked to avoid material failure. 
    
    \begin{figure*}[t]
    	\centering
    	\includegraphics[width=\textwidth]{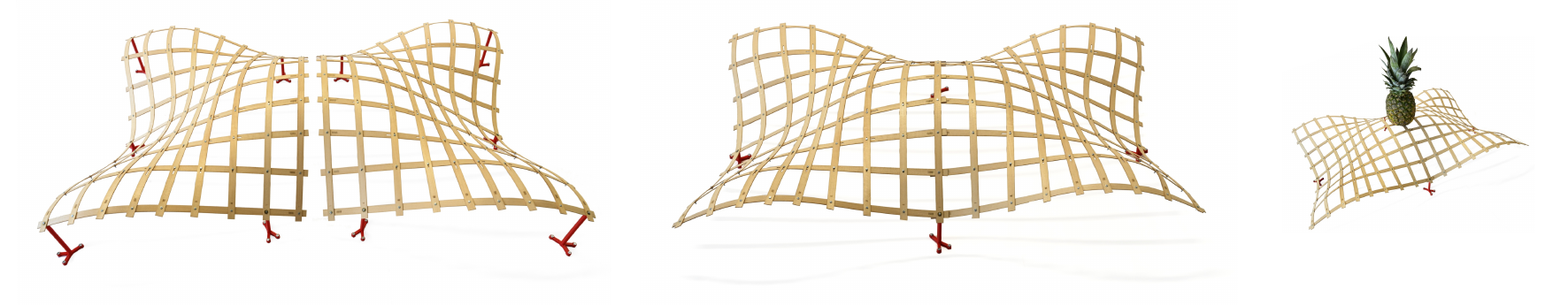}
    	\vspace{-10pt}
    	\caption{Modules of the small-scale prototype and tests on the structural behaviour. Left and Middle: Modules can be used separately or combined. Right: A test of the load-bearing capacity of the grid. }
    	\label{fig:small_scale}
    	\label{fig:structural}
    \end{figure*}

    \subsection{Deployment} \label{sec:deploy}
    In theory, an elastic grid can be deployed to match all isometries of the input surface patch. This is due to the nature of the design concept of elastic geodesic grids. If deployment is not guided, the grid will take the shape corresponding to minimal elastic energy. This shape will most certainly not correspond to the shape of the input surface patch.
    
    To encounter this, first, the deployment process of the grid needs to be guided.
    This means that the shape induced by deployment needs to be checked and adjusted, if necessary. It has to be ensured that the grid buckles in the right way. 
    To put it simply: It needs to be checked that all bumps that are supposed to go up actually go up. Adjusting the shape of the grid is easy in the beginning of deployment, as elastic forces are still rather low.
    Depending on the complexity of the design surface, however, this step can be tricky.
    
    Second, after having deployed the grid correctly,  it needs to be bent to match the shape of the surface patch. To fix the grid to this shape, a set of supports is used. Good choices for locations are intersections of grid curves on the surface patch. A support is defined by a point and a plane. The point corresponds to the intersection point of grid curves, and the plane corresponds to the tangent plane of this surface point.

    \section{Elastic Geodesic Grids Models}\label{sec:pavilion}
    
    For the design of an EGG with a couple of meters in size, first a small-scale prototype was built. It was intended to assess the aesthetic qualities of the design and provide insights for scaling.
    The first scaling-attempt failed, because of poor material properties. However, the second attempt was successful. The production process was automated as much as possible, e.g. by laser-cutting the notches and the PTFE-stickers.

	\subsection{Small-Scale Prototype}
	The prototype was designed by an architecture student without particular prior knowledge about elastic structures or differential geometry. The student was given remote access to a computer on the institute, where she generated the design surface in Rhinoceros. The EGG-pipeline could then be accessed via a Grasshopper node, and all computations happened in the background in \textsc{Matlab}.
	The simulated shape of the grid and the ready-to-use data for laser-cutting were output again in Grasshopper.
	
	Out of many design candidates we chose one that was well suited for splitting into modules (cf. Figure 3, one half of the symmetric surface patch).
    The measurements of the small-scale prototype were 0.85 x 0.57 x 0.15 meters, and it was built from 1 mm thick and 10 mm wide lime wood lamellas. A plan for laser-cutting the lamellas is provided in Figure \ref{fig:plans}. The supports have inclined contact areas and were 3D-printed. The prototype kept its shape even after being removed from the supports (cf. Figure \ref{fig:pla_dep}). As we wanted to keep the number of supports low, this was a relevant design feature.
    
    Tests also suggested a high degree of structural stiffness and load-bearing capacity.
    The total weight of the model was 160 grams, the applied weight in Figure \ref{fig:structural} was 1135 grams. This gives a promising load-to-self-weight ratio of about 7. 
    
    \begin{table}[b]
        \caption{The specific modulus $\lambda$ of the materials used ($\perp$ and $\parallel$ indicate fibre direction). Limewood was used for the small-scale prototype, the failed version was built from birch-plywood, and the successful medium-scale model was built from poplar-plywood.}
        \begin{tabular}{ccccc}
            &&Limewood& Plywood & Plywood   \\
            Parameter & Unit    &   $\parallel$ &birch $\perp$  & poplar  $\parallel$\\[0.05cm]
            \midrule
            $E$    &      [$GPa$]                   & 9.1   & 4.0   & 7.6            \\
            $\rho$ &[$\nicefrac{g}{cm^3}$]          & 0.50   & 0.65  & 0.43  \\
            $\lambda=\nicefrac{E}{\rho}$&[$10^{6}\cdot\nicefrac{m^2}{s^2}$]  & $18.2$ & $6.15$ &  $17.7$ 
        \end{tabular}
        
        \label{tab:specific}
    \end{table}
    
	\subsection{Fabrication Process}
	Fabricating the medium-scale EGG model consisted of the following steps: laser-cutting the lamellas and the notches, sanding and coating the lamellas, laser-cutting the PTFE-stickers, assembling the grid, and casting the supports. The different steps of the fabrication process can be seen in Figure \ref{fig:process}.
	
	As construction material plywood was used. It is cheap, available in big panels, and easy to machine. The plywood we used had three layers and was 3 mm thick. We decided to use a laser-cutter to produce the lamellas and notches.
	Our Trotec Speedy 500 has a back flap that can be opened. This enables easy production, as the plywood panels could be pushed through the flap. 
	The cutting happened in multiple steps: after cutting one segment, the plywood panel was pushed forward and adjusted to cut the next segment.  
    
    The supports were cast from concrete and designed to be stable under their own weight, but also light enough that they could be carried.
    The inclined contact areas of the supports were realized with wood wedges mounted to the supports.

	\subsection{Failed Version}
	
	\begin{figure}[b]
    	\begin{minipage}[b]{0.23\textwidth}
    	\centering
		\includegraphics[width=\textwidth]{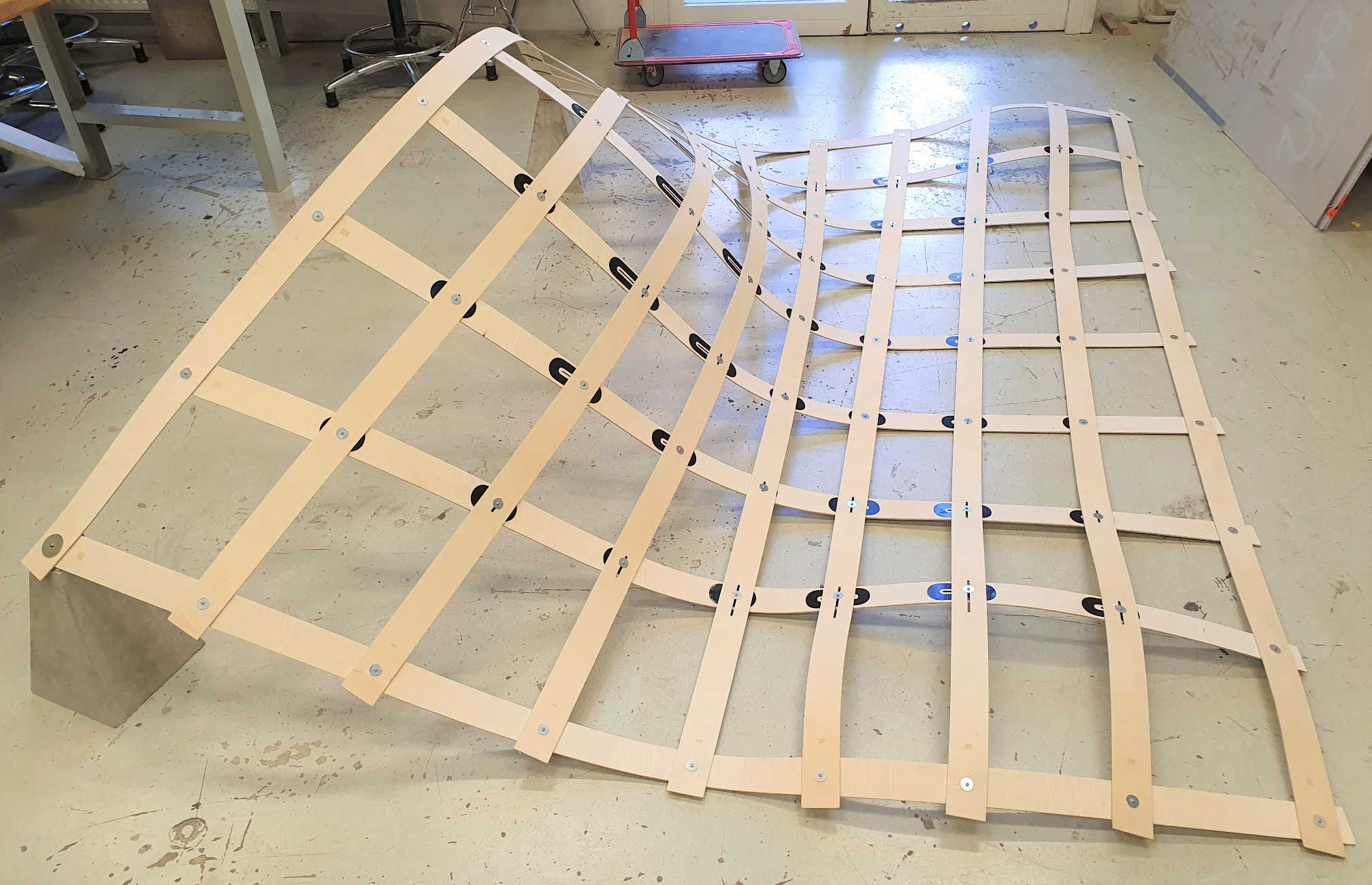}
	\end{minipage}
    	\hspace{0.6cm}
    	\begin{minipage}[b]{0.115\textwidth}
    	\centering
		\includegraphics[width=\textwidth]{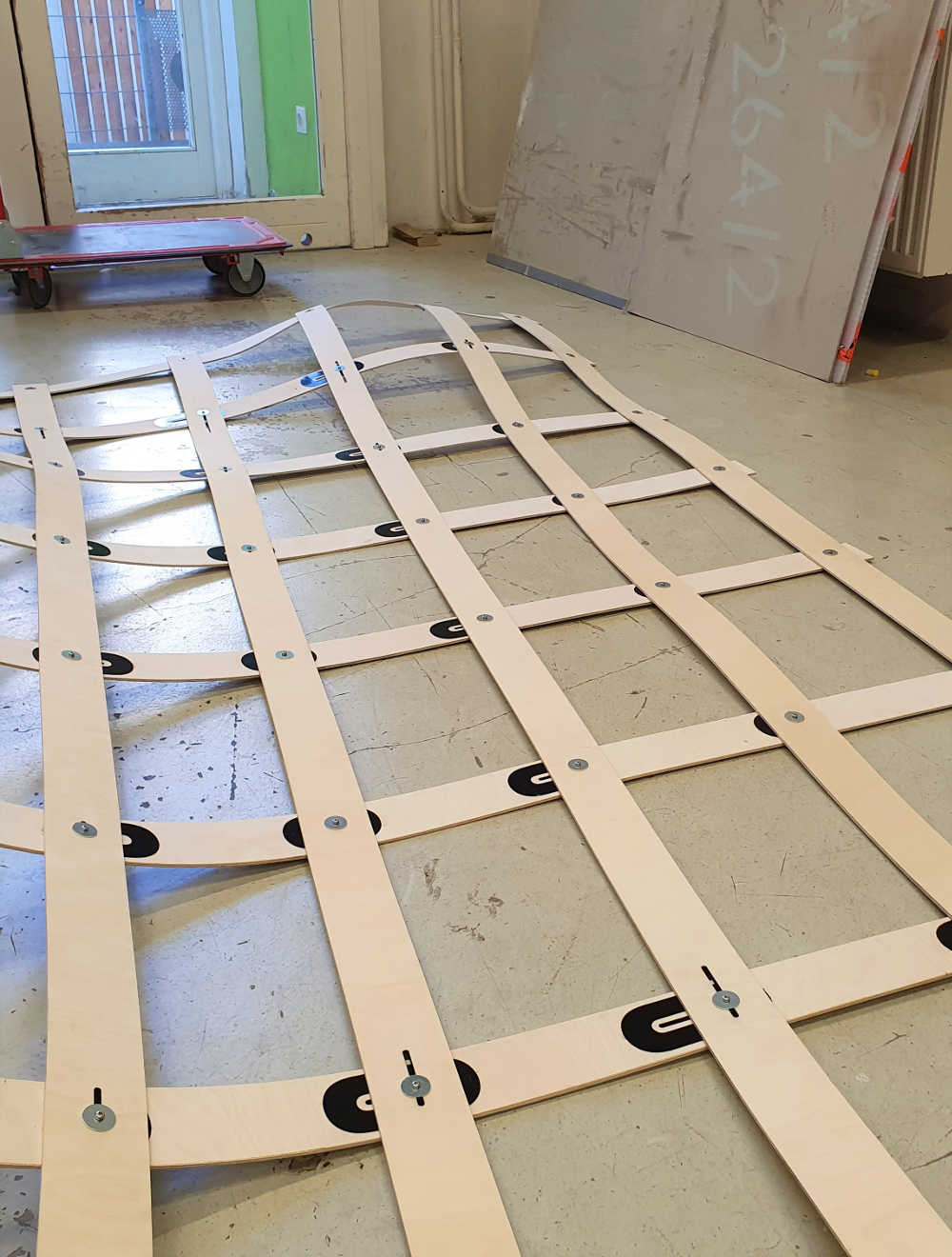}
    	\end{minipage}
    	\caption{One module of the failed medium-scale model. The structure is not able to carry its self-weight. Some curvature features can be recognized, however, the structure as a whole performs rather poorly.}
    	\label{fig:fail}
    \end{figure}

	In the first iteration of building the medium-scale model, the change of material from high-quality lime wood to simple plywood led to the failure of the structure (see Figure \ref{fig:fail}). 
	The reason for the failure was the wrong fibre-orientation of the top layers of the plywood. The delivered plywood panels were cut in an unfavorable way, which made it impossible to cut out the lamellas parallel to the fibre orientation. This made the lamellas excessively flexible, as the elastic modulus $E$ of timber perpendicular to the fibre direction is only about $\nicefrac{1}{5}$ of $E$ parallel to fibre direction. Essentially, the top layers hardly contributed to the structural performance, but made up $\nicefrac{2}{3}$ of the self-weight. We still tested if the material was usable, but unfortunately it was not.

	The insights on scaling in Section \ref{sec:scaling} explain this failure.
	Table \ref{tab:specific} summarizes the values for the specific modulus for the materials we used.
	When comparing the specific modulus of plywood (birch) to limewood, it can be seen that it is only about $\nicefrac{1}{3}$.
	Essentially, the plywood was way less performative and much heavier, which is exactly the opposite of what would have been appropriate for a larger grid.
	
	When scaling an elastic structure, the specific modulus $\lambda = \nicefrac{E}{\rho}$ should be a primary criterion for the choice of material.
	A second important criterion for the  choice of materials is the breaking stress over the elastic modulus, as proposed by \cite{gengnagel2013active}.

	\subsection{Medium-Scale EGG Model}
\begin{figure}
	\centering
	\includegraphics[width=\columnwidth]{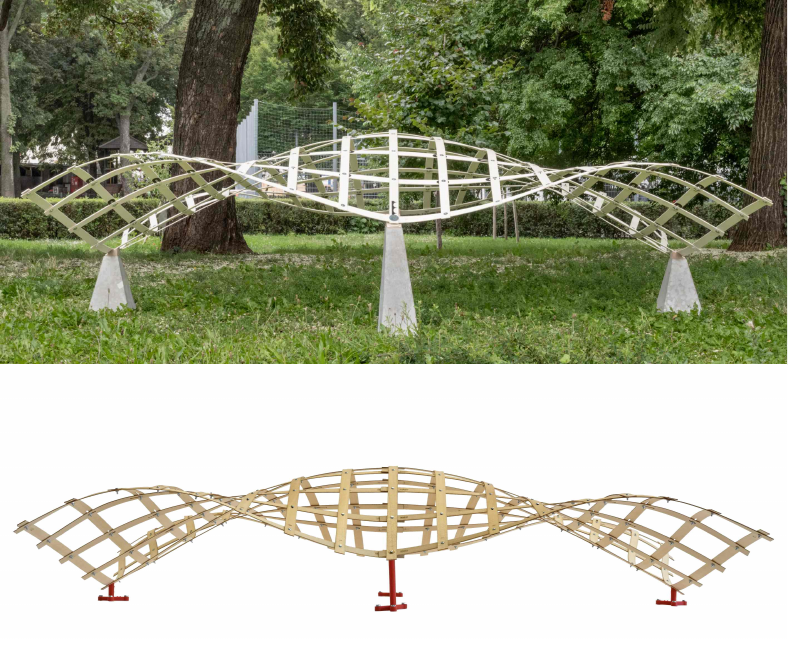}
	\vspace{-20pt}
	\caption{Side-view comparison of the shapes of the small-scale prototype and the medium-scale model. Good agreement of the shapes can be recognized.}
	\label{fig:side}
    \end{figure}

\begin{figure*}[t]
    \centering
	\includegraphics[width=\textwidth]{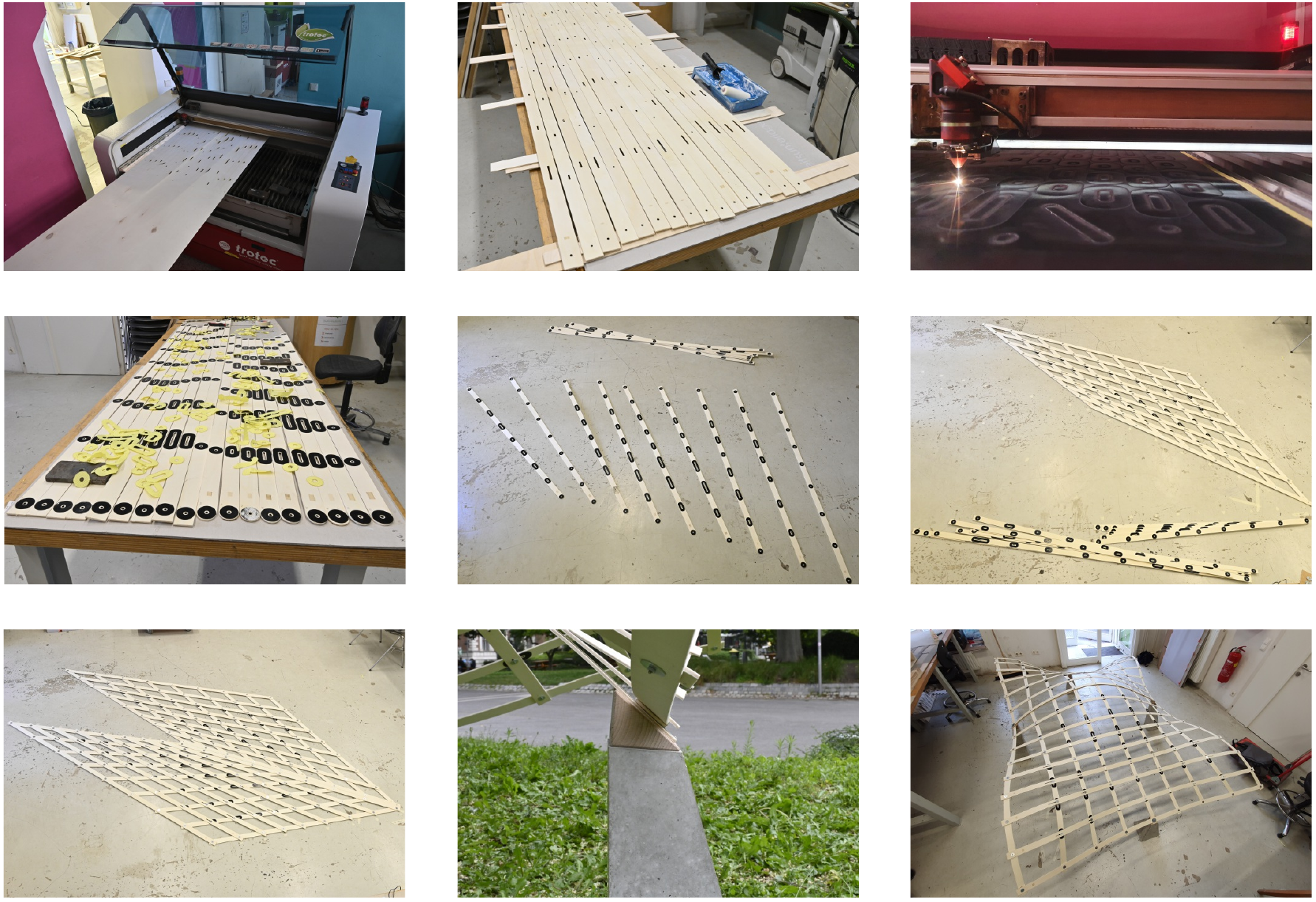}
	\caption{Some steps of the fabrication-process of the model. A laser-cutter was used to produce the lamellas and the PTFE-stickers. After sticking them to the lamellas, the modules of the grid were assembled. The supports were cast from concrete and have inclined contact areas. Finally the grid was deployed and fixed to the supports.}
	\label{fig:process}
    \end{figure*}

\begin{figure}
	\centering
	\includegraphics[width=\columnwidth]{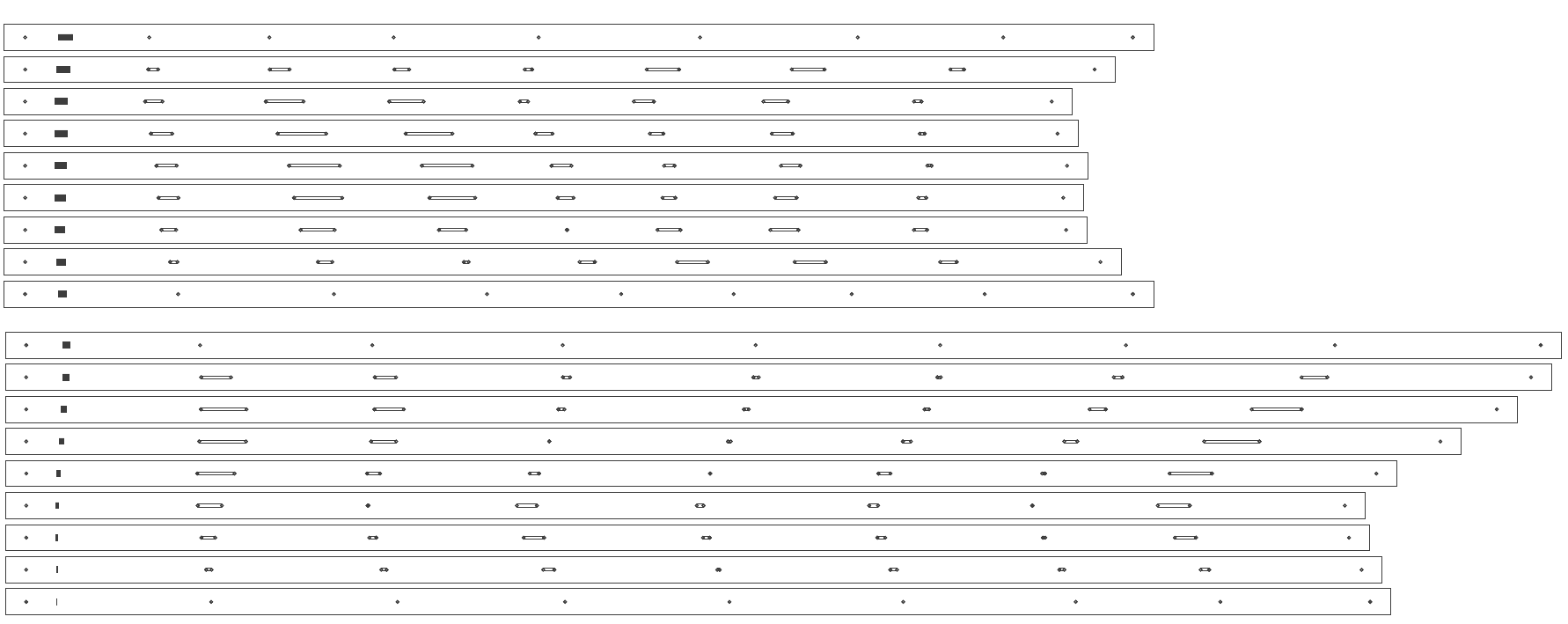}
	\caption{Plan for laser-cutting the lamellas. Scaling the picture size factor of 9 yields the size of the desktop model.}
	\label{fig:plans}
    \end{figure}

\begin{figure*}
	\centering
	\includegraphics[width=0.99\textwidth]{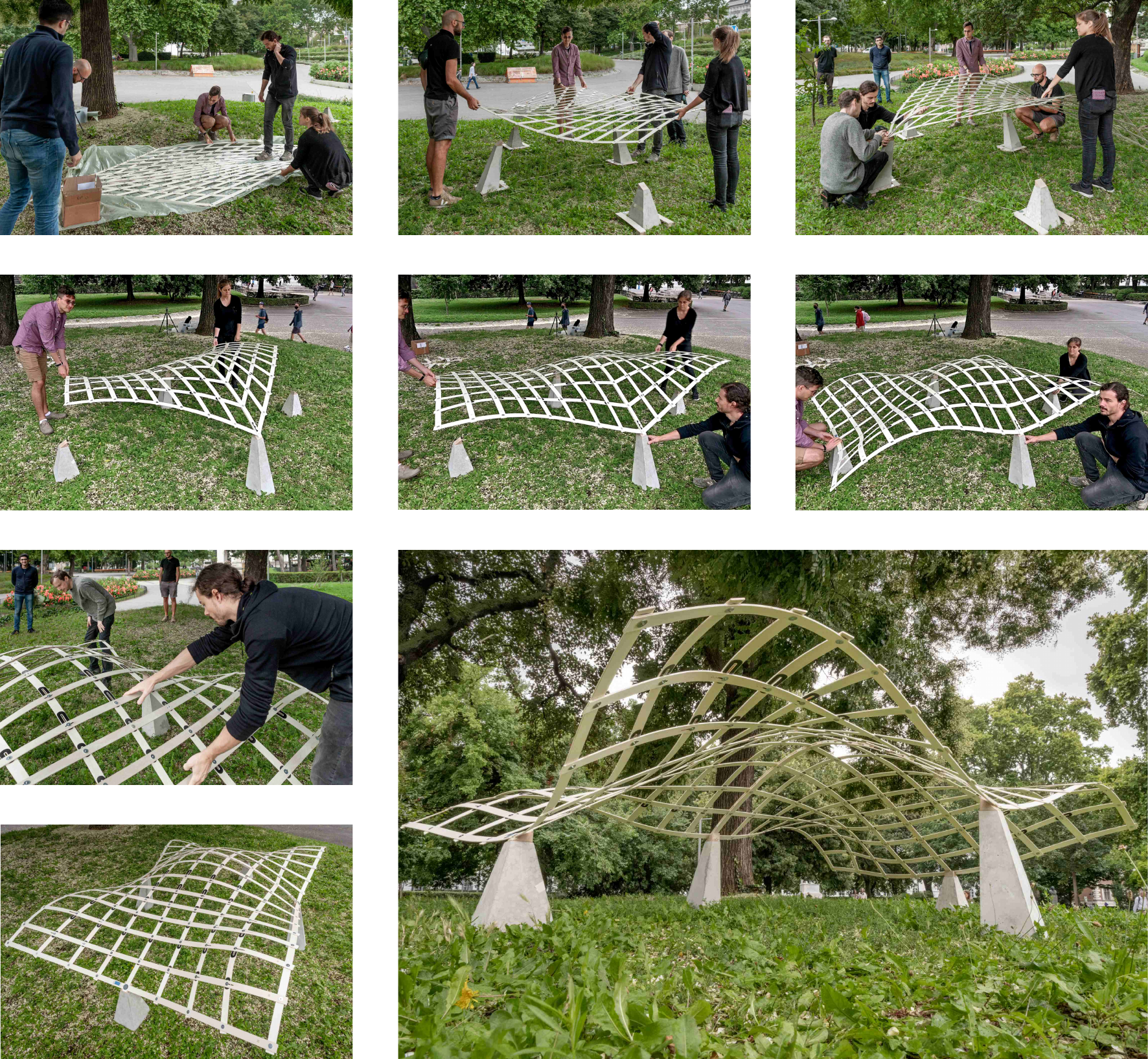}
	\caption{Several steps of the deployment of the EGG model and the deployed structure. The deployment process does not require a lot of force, and the curvature of the structure emerges naturally.}
	\label{fig:pavilion}
    \end{figure*}

	The second attempt to build the medium-scale model succeeded, using poplar plywood for the lamellas. This plywood is quite efficient, as Table \ref{tab:specific} shows. 
	In the deployed state  (cf. Figure \ref{fig:pavilion}) the structure measures 3.1 x 2.1 x 0.9 meters (including supports) and has a self weight of 7.1 kilograms. This makes a weight-to-span ratio of $1.09 ~ \nicefrac{kg}{m^2}$ and a thickness-to-span ratio of $\nicefrac{1}{516}$. In mathematical terms: with a structural thickness of 1 meter, 516 meters could be spanned (which is only possible in theory, of course). This, however, qualifies the model as an ultralight construction \cite{sobek}.
	To put this thickness-to-span ratio into context: The slender glass-steel construction of the Great Court of the British Museum by Foster and Partners has a thickness-to-span ratio of $\nicefrac{1}{200}$.
	The closeness of the shapes of the small-scale prototype and the medium-scale model is obvious in Figures \ref{fig:grids} and \ref{fig:side}.

	The deployment of the model was done by five people: four of them held the structure and one attached it to the supports. Some intermediate steps of the deployment process can be seen in Figure \ref{fig:pavilion}. 
	In Section \ref{sec:deploy} the challenges of deployment were discussed.
	On the site, the grid was first bent and fixed to two supports, then deployed. Pre-bending the grid before deployment eliminated problems with buckling into undesired configurations, and the grid automatically deployed correctly. Deployment worked smoothly and could be carried out by a single person. Sliding of the lamellas in the notches worked partially, however, the lamellas could be pushed into the right configuration by a single person without applying much force. 
	
	\section{Discussion and Conclusions}\label{sec:discussion}
	We presented a medium-scale EGG model of some meters in size to investigate the scalability of the elastic geodesic grids approach. Furthermore, we analyzed some design challenges that come with these special structures. Solutions to these challenges were proposed and tested on the structure. Among these, the necessity of smoothing the design surface has been avoided. More practical problems that were tackled are the manageability, the minimization of friction and strategies for successful scaling. We also presented a simple and fast fabrication process for the structure.
	
	The comparison of the shapes of the medium-scale model and the small-scale model shows satisfying closeness. 
	Some benchmarks for the structural performance were also presented.
	Investigating how the structural capabilities change with the change in size is out of scope of this paper. However, further research on this topic is an interesting path for future work.
	
	To reach architecturally relevant scales, the presented medium-scale model would have to be scaled even further. 
	However, insights into the mechanics of scaling and using high-performance, high-$\lambda$ materials suggest that further scaling is feasible.
	
	\begin{acks}
    This research was funded by the Vienna Science and Technology Fund (WWTF ICT15-082). 
\end{acks}
       
	\bibliographystyle{ACM-Reference-Format}
	\bibliography{papers-scf}

\end{document}